\pdfoutput=1
\documentclass[sigconf]{acmart}

\AtBeginDocument{%
  }

\copyrightyear{2026}
\acmYear{2026}
\setcopyright{rightsretained}
\acmISBN{979-8-4007-0985-2/24/08}

\usepackage{tikz}
\usetikzlibrary{positioning}
\usepackage[most]{tcolorbox}
\begin{document}

\title[FaulT-Bench]{FaulT-Bench: Towards Benchmarking Network Troubleshooting LLM Agents under Unreliable User Tickets}


\author{Kuan-Hao Tseng*}
\affiliation{%
  \institution{The University of Sydney}
  \country{Australia}}

\author{Niruth Bogahawatta*}
\affiliation{%
  \institution{The University of Sydney}
  \country{Australia}}

\author{Yasod Ginige*}
\affiliation{%
  \institution{The University of Sydney}
  \country{Australia}}

\author{Kunjan Patel*}
\affiliation{%
  \institution{The University of Sydney}
  \country{Australia}}

\author{Kosta Dakic*}
\affiliation{%
  \institution{The University of Sydney}
  \country{Australia}}

\author{Suranga Seneviratne*}
\affiliation{%
  \institution{The University of Sydney}
  \country{Australia}}

\renewcommand{\shortauthors}{Ginige et al.}
\newcommand{\complete}[1]{\textcolor{purple}{#1}}

\begin{abstract}
LLM-based agents are increasingly proposed for network fault diagnosis, but existing benchmarks evaluate them only on accurate tickets and always assume a fault is present, conditions rarely met in practice. We present FaulT-Bench, a benchmark of 200 troubleshooting scenarios across eight network topologies, five reimplemented from public practitioner labs, spanning genuine faults, false fault reports, incorrect device attribution, and incorrect root-cause claims. To isolate how ticket wording affects diagnosis, we further rewrite 72 false-premise tickets into five reporter personas that vary reporter confidence and verifiable detail one factor at a time, holding the network state fixed. Our automated harness deploys each scenario in Kathará, lets agents interact through the NIKA tool interface, and scores free-text diagnoses with an LLM judge across outcome, fix, and reasoning quality. Evaluating SADE, ReAct, and Claude Code, we find all three are near-saturated on accurate tickets and robust to misdirection, yet degrade sharply when the network is healthy and the ticket is wrong, probing until a benign condition can be promoted to a root cause rather than concluding nothing is wrong. Persona rewrites show that how a ticket is written matters more than what it claims: a confidently wrong report is handled about as well as an accurate one, while a vague, underspecified report degrades performance sharply. The three agents also fail differently, from constant over-diagnosis to unanswered runs, at very different cost. These results position FaulT-Bench as a benchmark for developing agentic systems that can reason reliably over the noisy, unreliable tickets of real-world network troubleshooting.
\end{abstract}





\maketitle
\section{Introduction}\label{sec:intro}

Modern computer networks are increasingly difficult to troubleshoot, and failures can cause widespread service disruptions with substantial economic and societal impact. High-profile incidents, including the 2021 Meta outage, in which a single BGP misconfiguration disrupted Meta's services worldwide for six hours, demonstrate how a minor configuration error can escalate into a large-scale outage~\cite{meta2021outage}. Recovering from such incidents is time-consuming because engineers must localize complex, multi-layer faults in large networks~\cite{meta2021outage,optus2024review,fcc2024att}. To reduce operational complexity, recent research has explored network management automation through intent-based networking and zero-touch service management~\cite{rfc9315,liyanage2022zsm}. More recently, Large Language Models (LLMs) have attracted significant attention and have been applied to a wide range of networking tasks~\cite{boateng2026survey,bariah2024telecom}, including configuration generation and validation~\cite{mondal2023hotnets,netconfeval2024}, network management automation~\cite{mani2023hotnets}, and root-cause recommendation from historical reports~\cite{ahmed2023icse}. However, diagnosing an ongoing fault in a live network remains a fundamentally different challenge. It requires iterative evidence collection, hypothesis refinement, and root-cause localization through direct interaction with the network. To tackle this challenge, recent work has developed LLM agents that iteratively diagnose such faults on emulated networks. In particular, NIKA~\cite{nika2025} deploys fault scenarios on the Kathará  emulator~\cite{kathara2018} and exposes diagnostic tools that troubleshooting agents, such as SADE~\cite{tseng2026sade}, can use to probe the network.


Despite recent progress, the evaluation of autonomous network troubleshooting agents remains limited. Existing benchmarks rely primarily on fault reports that accurately describe the underlying problem. They are less representative of real operational settings, where diagnosis begins from human-written trouble tickets that are frequently ambiguous, incomplete, inconsistent, or even misleading. To our knowledge, NIKA is the only publicly available benchmark for evaluating agentic troubleshooting on live emulated networks. While NIKA provides an important evaluation platform, it assumes that every scenario contains a genuine fault, and every diagnosis begins from a blind start without a user-written report. Consequently, it cannot assess whether an agent can verify that a reported fault actually exists, or distinguish a genuine failure from an erroneous user report. Evaluating this capability is essential before deploying autonomous troubleshooting agents in production, as minimizing false alarms is critical for efficient network operations. A false diagnosis can mislead engineers into repairing healthy devices while the real issue remains unresolved. Addressing these limitations requires benchmarks that better reflect operational environments by incorporating realistic, user-generated fault descriptions.

To this end, we present \emph{FaulT-Bench}, a benchmark for evaluating free-text diagnosis of emulated networks using realistic trouble tickets. The benchmark comprises 200 scenarios that present the user tickets for network issues across eight network topologies. 
Scenarios in FaulT-Bench can be divided into four categories: (i) \textit{Correct-Fault}, where a genuine network fault is injected, and the accompanying ticket correctly describes the observed symptoms; (ii) \textit{False-Premise}, where the network is healthy but the ticket incorrectly reports a fault; (iii) \textit{Wrong-Device}, where the ticket incorrectly attributes the fault to a healthy device; and (iv) \textit{Wrong-Cause}, where the reported cause is inconsistent with the actual root cause of the injected fault. To further evaluate how LLM agents react to human reporting styles, each ticket in the False-Premise scenarios, which we refer to as the \emph{standard} ticket, is rewritten in five reporter personas: a vague, uncertain \emph{novice user}, a confident but mistaken \emph{naive user}, and three controls that vary the reporter's confidence and the amount of verifiable detail in the ticket, one factor at a time. 


To facilitate LLM agent evaluation, FaulT-Bench provides an automated evaluation harness that deploys each scenario in the Kathará emulator, enables agents to interact with the network through the NIKA MCP tool, and assesses their free-text diagnoses using an LLM-as-a-judge framework across three dimensions: diagnostic outcome accuracy, repair recommendation quality, and reasoning quality. Finally, we evaluate three agents on our benchmark and report the results: SADE~\cite{tseng2026sade}, NIKA's ReAct agent~\cite{yao2023react,nika2025}, and Claude Code~\cite{Claude_code}. Overall, we show that while the agents do well on resolving Correct-Fault, Wrong-Device, and Wrong-Cause scenarios, their performance degrades substantially on False-Premise scenarios, frequently attributing faults to healthy devices. Closing the gap on unreliable, human-authored tickets is a necessary step before current agentic frameworks can offer a significant value proposition in production network troubleshooting, and, to the best of our knowledge, FaulT-Bench is the first benchmark to enable such research. Our contributions are as follows: 




\begin{itemize}
  \item We construct a benchmarking dataset of 200 free-text troubleshooting scenarios spanning eight network topologies: 80 \textit{Correct-Fault} scenarios, in which an injected fault is reported accurately by the tickets, and 120 incorrect-reporting scenarios, comprising 72 \textit{False-Premise}, 24 \textit{Wrong-Device}, and 24 \textit{Wrong-Cause} scenarios, to enable the evaluation of LLM agents under realistic user-reporting conditions. The dataset also contains 360 rewritten tickets, obtained by rewriting each of the 72 \textit{False-Premise} tickets in five reporter personas ($5 \times 72 = 360$), with controls that vary reporter confidence and the amount of verifiable detail. The dataset, the harness, and logs for every run are released.\footnote{\url{https://github.com/Overlxrd-uwu/FaulT-Bench}}
  \item We develop an automated evaluation harness that deploys network topologies in Kathará, verifies the injected fault, enables agent interaction through NIKA's diagnostic tools, and evaluates free-text diagnoses using an LLM judge across three dimensions: diagnostic outcome accuracy, repair recommendation quality, and reasoning quality. 
  
  \item We conduct an extensive evaluation of three troubleshooting agents. While all agents perform well on conventional fault scenarios and resist misdirection (Wrong-Device and Wrong-Cause shift accuracy by only $+2.5$ and $-2.1$ percentage points relative to the Correct-Fault mean), realistic tickets expose significant weaknesses. Agents falsely diagnose up to 24\% of healthy networks (SADE 24\% versus Claude Code 11\%), and ticket phrasing alone changes diagnostic accuracy by up to 14.5 percentage points, driven by how much of the report can be verified against the network rather than by how confidently it is written. 
\end{itemize}




\section{Related Work}\label{sec:related}


Early work on network management and troubleshooting automation includes zero-touch service management~\cite{etsi_zsm_2018}, intent-based networking~\cite{ibn_overview_minhas2024}, and static analysis tools~\cite{batfish}. Subsequent efforts explored machine learning-based approaches, including graph neural network~\cite{configreco, netgenius}. More recently, Large Language Models (LLMs) and LLM-based agents have been increasingly proposed for network configuration management and for open-ended troubleshooting, an area that remains comparatively underexplored, as prior automation efforts have focused primarily on provisioning and configuration rather than diagnosing faults in operational networks.


\subsection{LLM Agents for Network Troubleshooting}

Zero-shot LLMs as virtual system administrators on emulated topologies were initially researched~\cite{donadel2024llms}, and subsequent agentic systems structure diagnosis through ReAct-style tool loops~\cite{yao2023react} and procedural-based policies~\cite{tseng2026sade}. A parallel line applies LLM agents to root-cause analysis in cloud and microservice systems~\cite{chen2024rca, roy2024rca, wang2024rcagent, ahmed2023icse}. Industrial deployments have followed, including dialogue-based diagnosis at ByteDance~\cite{wang2024netassistant}, multi-agent event handling at Meta~\cite{sun2025confucius}, and fault localization and root-causing systems at Alibaba~\cite{zhang2025bian, ren2026xihe, xu2026aida}; recent surveys consolidate this rapidly growing space~\cite{boateng2026survey}. However, these systems are developed and evaluated on incidents in which a fault genuinely exists and the report describing it is accurate. The dual failure modes of inventing a fault in a healthy network and of being steered by a reporter who blames the wrong device or misstates the mechanism are never measured, despite false positives being a dominant operational cost~\cite{wang2024netassistant, xu2026aida}. 

\subsection{Network Emulation Environments}
Lightweight container-based network emulators, including Netkit \cite{pizzonia2008netkit}, Kathar\'a~\cite{kathara2018} and its scalable successor Megalos~\cite{scazzariello2020megalos}, and Containerlab~\cite{containerlab}, provide the reproducibility and programmatic fault injection that agent evaluation requires, whereas VM-based platforms such as GNS3 trade these properties for vendor-OS fidelity. However, the topologies on which agents are actually evaluated~\cite{nika2025, donadel2024llms, tseng2026sade} are synthetic networks designed by the benchmark authors themselves. Moreover, translating production configurations onto emulation substrates is non-trivial, as vendor CLIs, ACLs, and VLAN constructs do not map one-to-one onto FRR and Linux primitives. FaulT-Bench contributes five operational Cisco/GNS3 enterprise networks translated into Kathar\'a/FRR labs with verified topology preservation. Three unmodified NIKA reference topologies are retained as a control: an agent's performance drop on the enterprise networks can then be attributed to the complexity of real-world configurations rather than to artifacts introduced by our translation pipeline, since both sets run on the same Kathar\'a/FRR tooling. 

\subsection{Benchmarks and Datasets}
\label{sec:related_bench}

General AI agent benchmarks~\cite{liu2024agentbench, xi2024agentgym} contain no networking environments, while networking-specific benchmarks target the inverse, static problem of configuration synthesis~\cite{netconfeval2024, aykurt2024netllmbench, mondal2023hotnets, mani2023hotnets}, broad management tasks~\cite{yuan2025netpress}, telecom-domain troubleshooting~\cite{bariah2026telcoagent}, or cloud incident lifecycles~\cite{shetty2025aiopslab} rather than L2/L3 networks. Closest to our work, NIKA~\cite{nika2025} provides hundreds of curated incidents over Kathar\'a-emulated networks with MCP tool interfaces. However, these benchmarks share a structural assumption: a fault is always present, tasks arrive as structured incident specifications, and scoring reduces to matching agent output against ground-truth label masks. Such evaluation cannot, by construction, reward the answer ``nothing is wrong,'' nor probe robustness to vague or confidently wrong reporters. Moreover, the broader LLM-evaluation literature has established that abstaining on false-premise or unanswerable queries is an unsolved capability that scaling does not fix~\cite{kirichenko2025abstentionbench}, and that models defer to confidently wrong user framings~\cite{sharma2024sycophancy}; no networking benchmark operationalizes either failure mode. FaulT-Bench fills this gap with 170 free-text tickets over real-derived and reference topologies, a majority-counter-example design, and rubric-based LLM-judge scoring of free-form diagnoses.
\section{FaulT-Bench Benchmark}\label{sec:method}

In this section, we first describe the dataset, followed by the FaulT-Bench evaluation framework.

\begin{table}[b]
\footnotesize
\vspace{-2mm}
  \caption{Summary of the eight network topologies.}
  \label{tab:networks}
  \small
  \setlength{\tabcolsep}{4.5pt}
  \begin{tabular}{lrrl}
    \toprule
    Network & Devices & Routers & Protocols and features \\
    \midrule
    Kath1~\cite{jeremysitlab}       & 20 & 12 & OSPF, VRRP, DHCP, dual ISP, ACLs \\
    Kath2~\cite{katejay2019college} & 30 &  3 & RIP; web, FTP, DNS services \\
    Kath3~\cite{imsiddhant_proj1}   & 32 &  9 & RIP, VLANs, ACL firewall \\
    Kath4~\cite{imsiddhant_proj2}   & 28 &  9 & OSPF; SSH-managed switches \\
    Kath5~\cite{internetworks2024ospflab}                           & 15 & 13 & OSPF, RIP, EIGRP redistribution \\
    nika\_ospf~\cite{nika2025}      & 15 &  6 & OSPF, static routes \\
    nika\_bgp~\cite{nika2025}       &  4 &  2 & BGP \\
    nika\_clos~\cite{nika2025}      &  7 &  5 & BGP Clos fabric \\
    \bottomrule
  \end{tabular}
\end{table}

\subsection{Network Topologies and Ticket Types}\label{sec:method-dataset}

FaulT-Bench comprises 200 scenarios across eight network topologies. Five topologies are reimplemented from publicly available practitioner-built laboratory networks, including four enterprise and campus labs~\cite{jeremysitlab,katejay2019college,imsiddhant_proj1,imsiddhant_proj2} and one multi-protocol GNS3 lab~\cite{internetworks2024ospflab}, while the remaining three are adopted from the NIKA benchmark~\cite{nika2025}. The selected networks provide diverse configurations, sizes, and features as summarized in Table~\ref{tab:networks}. 


Each topology contributes 25 scenarios, comprising 10 correct-fault scenarios and 15 incorrect or misleading ticket scenarios. The 10 correct-fault scenarios provide the topology, injected fault, ticket, and ground truth. The 15 incorrect or misleading scenarios consist of nine false-premise scenarios, three wrong-device scenarios, and three wrong-cause scenarios, each designed to evaluate the agent's ability to identify inconsistencies between the reported issue and the underlying network state. Across the benchmark, these comprise 72 false-premise, 24 wrong-device, and 24 wrong-cause scenarios in total. Figure~\ref{fig:scenario} shows a sample scenario file.

\begin{figure}[t]
  \centering
  \tcbset{scnbox/.style={enhanced, boxrule=0.5pt, arc=2pt, left=5pt,
      right=5pt, top=3pt, bottom=3pt, toptitle=2pt, bottomtitle=2pt,
      fonttitle=\scriptsize\bfseries\sffamily, coltitle=white,
      fontupper=\scriptsize, width=\columnwidth, nobeforeafter}}
  \begin{tcolorbox}[scnbox, colframe=blue!45!black, colback=blue!3,
      colbacktitle=blue!45!black,
      title={SHOWN TO THE AGENT \; (together with the topology overview)}]
    \texttt{[PROMPT-TO-AGENT]}\; the free-text trouble ticket
  \end{tcolorbox}

  \vspace{1mm}
  \begin{tcolorbox}[scnbox, colframe=black!55, colback=black!2,
      colbacktitle=black!55,
      title={HIDDEN FROM THE AGENT \; (drives the framework and the judge)}]
    \texttt{Scenario:\ Topology:}\; identifier and the network to boot\\
    \texttt{[INJECTION]}\; commands creating the fault (empty if none)\\
    \texttt{[POST-INJECT-CHECK]}\; probe that must confirm the fault holds\\
    \texttt{[GROUND-TRUTH]}\; what is actually wrong, or that nothing is\\
    \texttt{[FIX]}\; repair specification (grades the fix score)\\
    \texttt{[SCORING-AXES]}\; must/bonus assertions (grade the outcome)\\
    \texttt{[DIAGNOSTIC-PROCESS]}\; reference walkthrough (documentation)\\
    \texttt{[NIKA-LABEL]}\; taxonomy label where one applies
  \end{tcolorbox}
  \caption{Anatomy of a scenario file. Only the ticket reaches the agent;
  every other block exists to inject, verify, and grade.}
  \label{fig:scenario}
\end{figure}

To evaluate the sensitivity of LLM agents to variations in reporting style, we rewrite each of the 72 false-premise tickets using five distinct reporter personas while keeping the underlying network state unchanged. Table~\ref{tab:dataset} summarizes the resulting dataset composition. The ID column reports the mean number of distinct verifiable identifiers per ticket, such as IP addresses, prefixes, device names, interfaces, and domain names. More verifiable information provides a clearer starting point for diagnosis, allowing agents to quickly narrow down potential causes. All tickets were manually authored and independently verified by three authors of the paper. The ticket statistics reported in this section, including identifier counts, are generated using the counting script released with the dataset. The following paragraphs detail each category. \\ \vspace{-3mm}

\begin{table}[b]
  \caption{Dataset composition: 200 core
  scenarios; the five persona-rewrite sets are a controlled variant layer. ID: mean distinct verifiable identifiers per
  ticket.}
  \label{tab:dataset}
  \footnotesize
  \setlength{\tabcolsep}{3pt}
  \begin{tabular}{lrcll}
    \toprule
    Category & \# & ID & Network state & Ticket characteristics \\
    \midrule
    Correct fault   & 80  & --- & injected fault   & accurately describes the fault \\
    False premise   & 72  & 3.2 & verified healthy & reports a nonexistent fault (standard) \\
    Wrong device    & 24  & --- & injected fault   & blames a healthy device \\
    Wrong cause     & 24  & --- & injected fault   & asserts the wrong cause \\
    \midrule
    Novice          & 72 & 1.9 & verified healthy & non-technical, uncertain; omits cause \\
    \quad confident & 72 & 1.9 & verified healthy & \emph{novice content; demanding tone} \\
    Naive           & 72 & 2.1 & verified healthy & non-technical, urgent; asserts a cause \\
    \quad unsure    & 72 & 2.1 & verified healthy & naive content; certainty removed \\
    \quad\quad no detail & 72 & 0.0 & verified healthy & \emph{unsure content; identifiers removed} \\
    \bottomrule
  \end{tabular}
\end{table}

\begin{figure*}[!t]
  \centering
  \tcbset{ticket/.style={enhanced,
      boxrule=0.5pt, arc=2pt, left=4pt, right=4pt, top=3pt, bottom=3pt,
      toptitle=2pt, bottomtitle=2pt,
      fonttitle=\scriptsize\bfseries\sffamily, coltitle=white,
      fontupper=\scriptsize, width=0.325\textwidth, nobeforeafter}}
  \begin{tcolorbox}[ticket, equal height group=ticketrow1,
      colframe=black!60, colback=black!2, colbacktitle=black!60,
      title={STANDARD: technical, suspects a cause}]
    Helpdesk ticket \#7071: Location C users (PC11--PC14,
    192.168.200.0/24) report their default gateway (192.168.200.1)
    ``drops out for a few seconds at random'' and suspect router4's LAN
    interface is flapping or overloaded. They want the flapping gateway
    fixed. Please confirm the fault.
  \end{tcolorbox}\hfill
  \begin{tcolorbox}[ticket, equal height group=ticketrow1,
      colframe=orange!50!black, colback=orange!6, colbacktitle=orange!50!black,
      title={NOVICE: non-technical, uncertain, drops the cause}]
    Helpdesk ticket \#7071: Hello, the network here in Location C keeps
    cutting out for a few seconds at a time, at random --- all four of our
    machines (PC11 to PC14) freeze at the same moment and then it all
    comes back. It's been happening on and off all day. Someone said the
    thing to mention is ``the gateway'' and gave me a number,
    192.168.200.1, but that's all I know, sorry. Could someone please
    work out what keeps cutting out and make it stay up? Thanks!
  \end{tcolorbox}\hfill
  \begin{tcolorbox}[ticket, equal height group=ticketrow1,
      colframe=orange!50!black, colback=white, colbacktitle=orange!50!black!75,
      title={NOVICE-CONFIDENT (control): demanding tone}]
    Helpdesk ticket \#7071: The network here in Location C keeps cutting
    out and it is disrupting the whole site. A few seconds at a time, at
    random --- all four of our machines (PC11 to PC14) freeze at exactly
    the same moment, then it all comes back. It has been happening on and
    off all day. The person before me said the thing to mention is ``the
    gateway'' and left the number 192.168.200.1 --- that is everything I
    have, and frankly it should be enough. Work out what keeps cutting out
    and make it stay up. We cannot run the site like this.
  \end{tcolorbox}

  \vspace{1.5mm}
  \begin{tcolorbox}[ticket, equal height group=ticketrow2,
      colframe=red!55!black, colback=red!4, colbacktitle=red!55!black,
      title={NAIVE: non-technical, urgent, asserts a definite cause}]
    Helpdesk ticket \#7071: Hi!! Our internet here in Location C keeps
    HICCUPING --- every so often the gateway thing (192.168.200.1) stops
    answering for a few seconds and then it's back, totally at random.
    All four of us (the PC11 to PC14 machines) get it at the same time.
    The router port our site hangs off is obviously failing --- it keeps
    dropping us and picking us back up!! Please fix or restart whatever
    does 192.168.200.1 and make the hiccups stop!! Thanks!!!
  \end{tcolorbox}\hfill
  \begin{tcolorbox}[ticket, equal height group=ticketrow2,
      colframe=red!55!black, colback=white, colbacktitle=red!55!black!75,
      title={NAIVE-UNSURE (control): certainty removed}]
    Helpdesk ticket \#7071: Hi!! Our internet here in Location C keeps
    HICCUPING --- every so often the gateway thing at 192.168.200.1 stops
    answering for a few seconds and then it's back, totally at random.
    All four of us, the PC11 to PC14 machines, get it at the same time.
    Something around the router port our site hangs off must be going
    wrong, but I don't know what. Could someone look at whatever does
    192.168.200.1 and make the hiccups stop!! Thanks!!!
  \end{tcolorbox}\hfill
  \begin{tcolorbox}[ticket, equal height group=ticketrow2,
      colframe=red!55!black, colback=white, colbacktitle=red!55!black!75,
      title={NAIVE-NO-DETAIL (control): identifiers removed}]
    Helpdesk ticket \#7071: Hi!! Our internet here keeps HICCUPING --- every
    so often the gateway thing stops answering for a few seconds and then
    it's back, totally at random. All four of us get it at the same time.
    Something around what our site connects to must be going wrong, but I
    don't know what. Could someone look at whatever handles the gateway and
    make the hiccups stop!! Thanks!!!
  \end{tcolorbox}

  \vspace{1.5mm}
  \begin{tcolorbox}[enhanced, width=\textwidth, boxrule=0.5pt, arc=2pt,
      colframe=green!35!black, colback=green!4, left=5pt, right=5pt,
      top=3pt, bottom=3pt, fontupper=\scriptsize, nobeforeafter]
    \textbf{Ground truth (identical for all six):} the network is
    verified healthy; PC11 reaches 192.168.200.1 with 0\% loss and the
    accused interface is stable. Correct diagnosis:
    \texttt{is\_anomaly:\;false}.
  \end{tcolorbox}
  \caption{ Six ticket forms of one false-premise scenario: the
  \emph{standard} ticket and the novice pair (top), the \emph{naive}
  chain (bottom). Each control changes one aspect relative to its parent, while the network and the ground truth remain the same.}
  \label{fig:ticket-example}
\end{figure*}

\noindent{\textbf{Correct fault:}} 
In correct-fault scenarios, we inject a real fault into the network and
give the agent a ticket that describes the resulting symptoms accurately.
The injected faults span service and daemon failures, configuration
errors, link failures, and filtering faults. For example, one scenario
stops the DNS service on a server host, and the ticket reports that users
can no longer reach sites by name.
These scenarios measure the agent's baseline ability to localize and
explain a genuine fault from an accurate report, which is the setting
that existing benchmarks test, and they anchor the comparison for the
misleading categories. Each topology contributes ten correct fault scenarios, resulting in 80 scenarios across the eight networks. \\ \vspace{-3mm}

\noindent{\textbf{False premise:}}
In false-premise scenarios, no fault is injected, and the network remains in its verified-healthy baseline state while the ticket reports a non-existent fault, such as a failed service, flapping route, or unreachable location. For example, as shown in Figure~\ref{fig:ticket-example}, the ticket reports a flapping gateway even though the gateway is stable. These scenarios evaluate whether the agent verifies the reported issue against the live network rather than blindly accepting the ticket premise. To ensure reliable evaluation, the relevant baseline behavior is verified beforehand, and intentional network policies, such as inter-office access restrictions, are documented to avoid incorrectly treating them as faults. Each topology contributes nine scenarios, resulting in 72 scenarios in total. We call these tickets \textit{standard} before the rewriting in different reporting styles.\\ \vspace{-3mm}

\noindent{\textbf{Wrong device:}}
In wrong-device scenarios, a real fault is injected; however, the ticket
confidently points to a different, correctly functioning device. For example, a ticket may identify an access switch as the source of the problem when the actual fault is a stopped routing daemon on an upstream router. These scenarios assess whether the agent can avoid such misdirection by ruling out the incorrectly accused device and correctly localizing the underlying fault. Each topology contributes three scenarios, resulting a total of 24. \\ \vspace{-3mm}

\noindent{\textbf{Wrong cause:}}
In wrong-cause scenarios, a real fault is injected, and the ticket points
near it but asserts an incorrect mechanism. For example, an outage may be reported as resulting from a failed link when the actual cause is a missing route. These scenarios assess whether the agent can distinguish between the observed symptoms and the proposed explanation, reject the incorrect hypothesis, and identify the actual root cause. Each topology contributes three scenarios, resulting in 24 scenarios in total.\\

\subsection{Tickets with Different Reporting Personas}\label{subsec:persona_rewrites}

To evaluate the sensitivity of LLM agents
to variations in reporting style, we rewrite each false-premise ticket
using two reporter personas and three control variants. All rewritten tickets run on the
same verified-healthy networks, ensuring that the underlying network
state remains identical across personas and that differences in agent
behavior can be attributed to the wording of the ticket rather than to
the correctness of the reported fault. We manually authored all tickets, and three authors independently verified them.

The two personas describe the same symptom in non-technical
language while differing in how they express the reporter's theory. A
{\bf \textit{novice}} reporter omits the suspected cause, uses uncertain and
apologetic language, and provides less verifiable detail. In contrast, a
{\bf \textit{naive}} reporter keeps the suspected cause, expresses it with high
confidence, and often prescribes a specific repair. 


Because each persona changes multiple aspects of the ticket simultaneously, we introduce controlled variants to isolate individual factors. Each persona is therefore paired with a corresponding control ticket, while the underlying network, injected fault, ground truth, and scoring rubric remain identical.
The {\bf \textit{novice-confident}}  variant modifies the novice ticket by replacing its apologetic and hesitant tone with a confident and demanding tone, while keeping all verifiable information unchanged. Specifically, every address, prefix, device name, interface, and domain is preserved, with no identifiers added or removed. These 72 matched pairs therefore isolate the effect of reporter confidence.
Similarly, the {\bf \textit{naive-unsure}} variant retains the naive reporter's description, urgency, and suspected cause but removes the reporter's certainty. The reporter acknowledges uncertainty (e.g., ``but I don't know what exactly''), avoids explicit certainty markers, and refrains from prescribing a specific repair. To further isolate the effect of verifiable detail, the {\bf \textit{naive-no-detail}} member of the pair removes the identifiers present in the naive ticket. Thus, the naive-unsure pair differs only in the presence or absence of verifiable identifiers after the reporter's certainty has been removed. This separation is important because removing identifiers while retaining a highly confident accusation would simultaneously weaken the accusation itself. Figure~\ref{fig:ticket-example} shows one scenario in its \emph{standard},
\emph{novice}, and \emph{naive} forms.


\subsection{Evaluation Framework}\label{sec:method-overview}

\begin{figure}[h!]
    \centering
    \includegraphics[width=\linewidth]{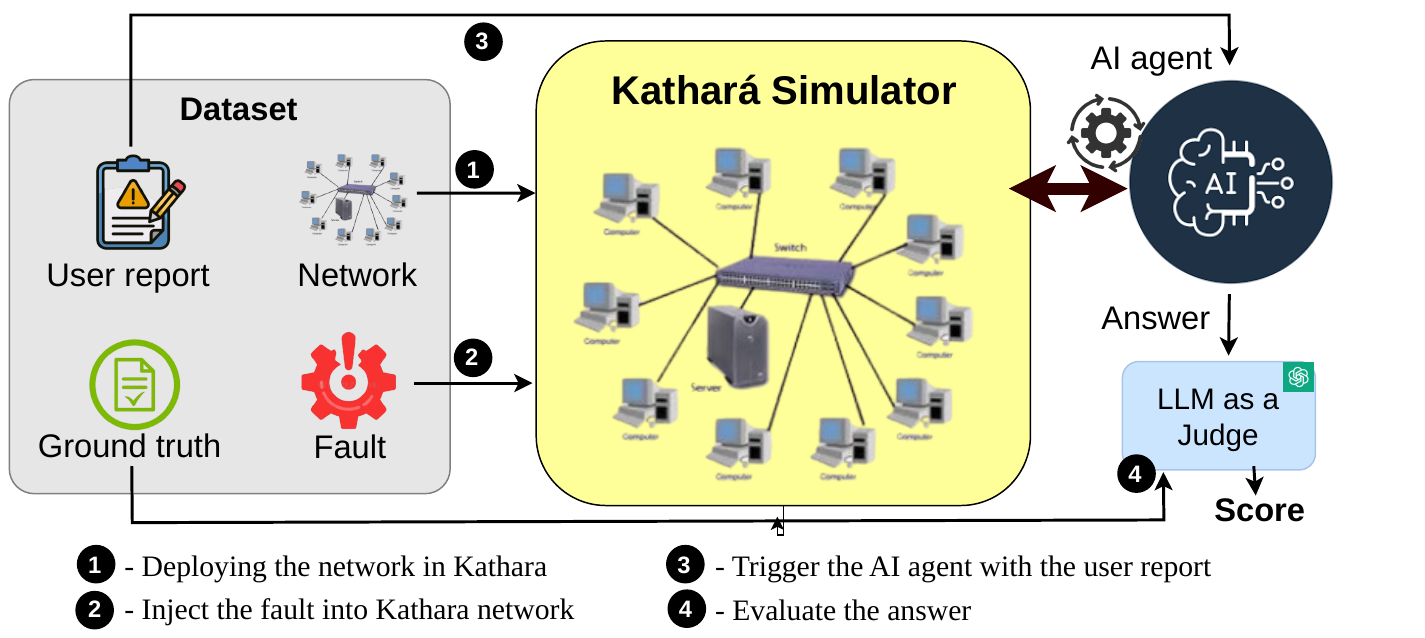}
    \caption{System overview of FaulT-Bench.}
    \label{fig:overview}
    \vspace{-3mm}
\end{figure}

In addition to the ticket dataset, FaulT-Bench provides an accompanying framework for assessing LLM-agent-based systems developed for network troubleshooting. The framework enables researchers to integrate their agents and automatically evaluate them.


Each scenario in FaulT-Bench can be automatically evaluated on an emulated network environment built with Kathará~\cite{kathara2018}, a container-based network emulator. Here, each network device is instantiated as an independent Docker container and interconnected according to the specified topology. Routers run the FRRouting, while end hosts operate standard Linux. Consequently, network behavior is produced by real networking software rather than by an abstract simulation: routing decisions are made by the FRR daemons, packet forwarding is handled by the Linux kernel networking stack, and injected faults affect the running system directly. Note that only the underlying hardware, physical links, and network topology are virtualized, allowing agents to interact with a network that reflects the actual execution state of its deployed components.

To enable controlled interaction with the emulated network, we adapt the NIKA tool interface~\cite{nika2025} and provide the agent with 22 tools. These tools support both host-level and router-level operations, including connectivity tests, interface and routing-table inspection, service queries, shell command execution, and network telemetry collection. The agent is provided with high-level topology information at the beginning of each scenario, including device, interface, and connectivity information. However, it does not have access to the underlying network configuration files (i.e., the Kathará files). The agent must therefore obtain configuration-level information required for diagnosis by actively probing the running network through the available tools. Each FaulT-Bench scenario is represented as a plain-text file containing multiple blocks with different visibility levels, as illustrated in Figure~\ref{fig:scenario}. The agent is provided only with the ticket, while the remaining blocks are used internally for fault injection, scenario verification, and evaluation.


Figure~\ref{fig:overview} illustrates the experimental workflow used to evaluate the AI agents. Each experiment follows four main steps. First, the network topology specified by the dataset is deployed in the Kathar'a simulator. Second, when the scenario specifies a fault, the corresponding fault is injected into the deployed network; for scenarios without a fault, the network remains in its baseline state. Third, the AI agent is triggered with the user report and interacts with the simulated network to investigate the reported issue and produce a diagnosis. Finally, the agent's answer is evaluated against the ground-truth information using an LLM-based judge. \\ \vspace{-3mm}

\noindent{\bf Network deployment:} For each scenario, the network topology provided in the dataset is instantiated in Kathar\'{a}, creating the devices and their network connections. The simulator provides an isolated and reproducible environment for the agent to investigate. \\ \vspace{-3mm}

\noindent{\bf Fault injection:} For scenarios containing a fault, the specified fault is introduced into the deployed network according to the ground-truth scenario. Following the injection, the network is actively verified to ensure that the fault has been successfully applied and that the resulting network state matches the expected ground-truth condition, using the scenario's [POST-INJECT-CHECK] block (Figure 2): its commands are run on the named devices through Kathará, and their output decides whether the injection succeeded. \\ \vspace{-3mm}

\noindent{\bf Agent interaction:} Once the network has been prepared, the user report is provided to the AI agent, triggering the diagnosis process. The agent interprets the reported symptoms and investigates the simulated network to identify the underlying cause. During this process, the agent can interact with the simulated network through NIKA tool calls. The collected observations are used to iteratively refine the diagnosis, after which the agent produces a final answer describing its findings and, if applicable, the identified root cause.

\subsection{Performance Metrics}\label{sec:method-performance}

After investigating the network, the agent must submit a structured diagnosis containing four elements: whether an anomaly exists (\textit{is\_anomaly}), the anomalous device (\textit{faulty\_devices}), an explanation of the root cause, and a proposed fix. The submitted diagnosis is then evaluated by an LLM-based judge using the ground-truth information and the evaluation criteria defined for the scenario.

As shown in Figure~\ref{fig:judge-example}, the evaluation is divided into three independent scores, each measuring a different aspect of the agent's diagnosis. \emph{Outcome score} measures the diagnosis accuracy and the explanation quality. The \emph{Fix score} evaluates the effectiveness and appropriateness of the proposed remediation steps. Finally, the \emph{Reasoning score} measures the quality of the reasoning process underlying the overall diagnosis. Note that the \emph{Fix score} and the \emph{Reasoning score} are two additional evaluation metrics introduced in FaulT-Bench compared to the NIKA dataset. As mentioned in Section~\ref{sec:related_bench} NIKA scores a diagnosis as a match against its ground-truth labels (anomaly, faulty components, root-cause label) and does not evaluate mitigation. Therefore, it can not credit a repair that would work or a conclusion that is actually supported by the evidence collected. The Fix and Reasoning scores add these two further evaluation angles to our benchmark.  For each metric, we feed the ground-truth block and the agent's generated final diagnosis to the same LLM judge (\texttt{gpt-5-mini}) using fixed prompts. These prompts are provided in our anonymous repository. The judge evaluates the diagnosis independently for each metric.  \\ \vspace{-3mm}

\noindent{\textbf{Outcome score:}} The Outcome score evaluates both the accuracy of the agent's diagnosis and the quality of its explanation. Diagnostic accuracy is determined by comparing the agent-identified faulty device and root cause against the ground truth. The quality of the generated explanation is assessed by an LLM judge, which compares the agent's explanation with the corresponding ground-truth explanation. The final Outcome score assigns a weight of 0.7 to diagnostic accuracy and 0.3 to explanation quality (i.e., $\mathrm{Outcome\_score} = 0.7\, \mathrm{diagnosis\_accuracy} + 0.3\, \mathrm{explanation\_quality}$). This weighting keeps correctness dominant: a correct diagnosis (at least 0.7) always outranks an incorrect one, however well explained (at most 0.3). \\ \vspace{-3mm}

\noindent{\textbf{Fix score:}} is calculated only for scenarios that contain a \texttt{[FIX]} block, namely the 80 correct-fault scenarios and the 48 wrong-device and wrong-cause scenarios. Healthy-network scenarios do not receive a Fix score because there is no fault requiring remediation. For each applicable scenario, the LLM judge is provided with the expected remediation and the agent's diagnosis and evaluates only the remediation proposed by the agent. The assessment is based on the substance of the proposed fix rather than its exact wording. A score of 1 is assigned when the proposed remediation would correctly resolve the fault, 0.5 when it is directionally correct but incomplete, and 0 when no remediation is proposed or the proposed remediation is incorrect. Thus, $\mathrm{Fix\;Score}\in{0, 0.5, 1}$. \\ \vspace{-3mm}

\noindent{\textbf{Reasoning score:}} The Reasoning score assesses the quality and completeness of the agent's diagnostic reasoning across three dimensions: \emph{grounding}, \emph{causality}, and \emph{coverage}. Following G-Eval~\cite{G-eval}, the LLM judge first provides a brief analysis and then assigns a score from 0--10 for each dimension using an anchored scale, where 0 indicates that the criterion is absent or contradicted, 3 indicates weak performance, 5 indicates partial satisfaction, 8 indicates mostly solid performance, and 10 indicates that the criterion is fully demonstrated. \emph{Grounding} measures whether the key claims in the diagnosis are supported by observations actually collected by the agent. \emph{Causality} assesses whether the agent's inferences logically follow from the available evidence without unsupported leaps or contradictions. \emph{Coverage} measures whether the diagnosis includes the essential inferences and correctly distinguishes the actual root cause from plausible alternatives.
To assess grounding, the judge is provided not only with the ground truth and the agent's final diagnosis, but also with a summary of the agent's investigation log, including all tool calls, truncated tool outputs, and intermediate notes. Consequently, a claim is considered grounded if the investigation log contains the supporting observation, even when that observation is not explicitly repeated in the final diagnosis. The final Reasoning score is calculated as $\mathrm{Reasoning} = (\mathrm{grounding} + \mathrm{causality} + \mathrm{coverage}) / 30 \in [0,1]$.


The judge is explicitly instructed to evaluate the validity of the reasoning rather than adherence to a particular diagnostic procedure. Thus, an agent that reaches the correct conclusion through a different but valid reasoning path can receive full marks. These scores evaluate different aspects of performance rather than requiring the agent to follow a predetermined diagnostic procedure. An agent may use any valid sequence of NIKA tool calls to investigate the network, provided that its final diagnosis is supported by the evidence it collected. The LLM judge generates the three scores by comparing the agent's final diagnosis with the ground-truth scenario and its corresponding evaluation rubric. If the agent fails to provide a final diagnosis, it is considered a timeout.

\begin{figure}[t]
  \centering
  \tcbset{jbox/.style={enhanced, boxrule=0.5pt, arc=2pt, left=5pt, right=5pt,
      top=3pt, bottom=3pt, toptitle=2pt, bottomtitle=2pt,
      fonttitle=\scriptsize\bfseries\sffamily, coltitle=white,
      fontupper=\scriptsize, width=\columnwidth, nobeforeafter}}
  \begin{tcolorbox}[jbox, colframe=black!60, colback=black!2, colbacktitle=black!60,
      title={SCENARIO \texttt{kath2-pc1-wrong-netmask} (hidden from the agent)}]
    \texttt{[INJECTION]}\; pc1's mask changed from /8 to /30, so its gateway 1.0.0.1 falls outside its subnet\\
    \texttt{[SCORING-AXES]}\; \texttt{assert-identifies: pc1}; \texttt{must-identify-device: pc1};
    \texttt{must-identify-component: netmask}; \texttt{bonus-explains-gateway-outside-subnet}\\
    \texttt{[FIX]}\; restore pc1's /8 mask and its default route via 1.0.0.1
  \end{tcolorbox}
  \vspace{1mm}
  \begin{tcolorbox}[jbox, colframe=blue!45!black, colback=blue!3, colbacktitle=blue!45!black,
      title={AGENT DIAGNOSIS (CC-Baseline, one run)}]
    \texttt{is\_anomaly: true}, \texttt{faulty\_devices: [pc1]}.
    ``pc1's eth0 has a /30 prefix instead of the /8 of the gateway; the kernel installs only the
    1.0.0.4/30 route, so 1.0.0.1 is unreachable \ldots\ reconfigure eth0 to 1.0.0.5/8 and add the
    default route via 1.0.0.1.''
  \end{tcolorbox}
  \vspace{1mm}
  \begin{tcolorbox}[jbox, colframe=green!35!black, colback=green!4, colbacktitle=green!35!black,
      title={JUDGE}]
    \textbf{Outcome:} 3 of 3 must/assert axes passed, 1 of 1 bonus axis passed:
    $0.7\cdot\frac{3}{3} + 0.3\cdot\frac{1}{1} = 1.0$ (0.7 had the bonus failed).\\
    \textbf{Fix:} the proposed remediation restores the mask and the default route and matches \texttt{[FIX]}: 1.0.\\
    \textbf{Reasoning:} grounding 8, causality 9, coverage 8 (the log shows \texttt{ip addr}, \texttt{ip route} and the failed ping on pc1): $(8+9+8)/30 = 0.83$.
  \end{tcolorbox}
  \caption{One run through the three scores. The judge sees the ground truth, the diagnosis, and for the reasoning score, a digest of the investigation log.}
  \label{fig:judge-example}
\end{figure}

\section{Baseline Agent Evaluation on FaulT-Bench}\label{sec:experiments}

 To demonstrate the feasibility and usefulness of FaulT-Bench for evaluating agentic frameworks in network troubleshooting, we evaluate three state-of-the-art agents, SADE, NIKA's ReAct agent, and Claude Code, spanning distinct agent design paradigms. Our evaluation asks three questions. {\bf RQ1} - How well do current LLM
agents localize and explain real faults on realistic networks when the ticket is accurate?,  {\bf RQ2} - Faced with a false-premise, wrong-device, or wrong-cause ticket, do agents correct the error or inherit it?, and  {\bf RQ3} - Holding information
content fixed, does the way a ticket is written change the diagnostic
outcomes?

\begin{table}[t]
  \caption{Evaluated agents: Access the same 22-tool
  interface, receive identical inputs, and have the same 20-turn
  budget.}
  \label{tab:agents}
  \footnotesize
  \setlength{\tabcolsep}{3.5pt}
  \begin{tabular}{llll}
    \toprule
    Agent & Base model & Beyond the 22 tools & Pairing isolates \\
    \midrule
    SADE~\cite{tseng2026sade} & \texttt{claude-sonnet-4-6} & skills + helpers & --- \\
    CC-Baseline~\cite{Claude_code} & \texttt{claude-sonnet-4-6} & nothing & skills layer \\
    ReAct~\cite{yao2023react,nika2025} & \texttt{gpt-5} & nothing & model + loop \\
    \bottomrule
  \end{tabular}
  \vspace{-3mm}
\end{table}

Table~\ref{tab:agents} lists the three agents. \emph{SADE}~\cite{tseng2026sade} is a skills-augmented diagnostic agent. Its fault-family playbooks and
lab-side helper scripts are what distinguish it, and they grant no
capability beyond compositions of the shared tools. \emph{CC-Baseline} is
Claude Code~\cite{Claude_code} itself, the same underlying model with the
same 22 tools. \emph{ReAct} is a classic
reason-and-act loop on a different base model over the identical tool
interface. All three receive byte-identical input (the
ticket plus the device-level topology overview of
Section~\ref{sec:method-overview}). Two experiment settings cover these three research questions. \\ \vspace{-3mm}

\noindent{\bf E1: Core benchmark (RQ1, RQ2).} Every agent runs on all 200
core scenarios in three independent passes, giving 1{,}800 runs. The 80
correct-fault scenarios answer RQ1. The 120 incorrect-reporting scenarios answer RQ2, and the contrast
between the two exposes failure modes that accurate-ticket evaluation
misses. Individual runs are noisy, with 8\% of (scenario, agent) cells
moving by 0.5 or more between passes, so every E1 figure is a three-pass
mean. \\ \vspace{-3mm}

\noindent{\bf E2: Persona rewrites (RQ3).} Every agent runs on the five
persona rewrites of the 72 false-premise tickets. Each rewrite is paired
with its standard-ticket counterpart from E1 over the same network state,
so a within-pair score difference is attributable to the manipulated
factor alone. Each persona ticket cateroy is run twice; the standard row it is
compared with comes from the three E1 passes. Every run follows the pipeline of Figure~\ref{fig:overview} on a freshly
deployed network. Runs execute serially,
one emulated network at a time, on a single commodity workstation; a run
takes about 6 minutes on average of wall-clock time, and a full pass over the core
benchmark takes $\sim$59 hours and costs $\sim$US\$165 of API usage.

All runs are graded by the same LLM judge as described in Section~\ref{sec:method-overview}. A run in which the agent terminates without
a parseable verdict is counted as a timeout and reported separately. 

\section{Results}
\label{sec:results}
 

Next, we present the results. All tables report means over the runs that answered, averaged over passes (three for the core scenarios, two for the persona rewrites). Due to space constraints, we report only the average values here. Per-pass values and standard deviations are reported in our artifact repository. In results tables \textit{Over-d.} is the share of answered runs on a healthy network that report a fault, and \textit{T.O.} is the share of runs that used up the turn budget without submitting a diagnosis, counted separately and never folded into score means; $\Delta$ is the difference from the Correct-Fault mean; CC-B.\ abbreviates Claude Code-Baseline.

\begin{table}[t]
\caption{E1: outcome by class and agent on core scenarios.} 
\label{tab:e1}
\footnotesize
\setlength{\tabcolsep}{3.5pt}
\begin{tabular}{@{}l cccc c cc@{}}
\toprule
& \multicolumn{4}{c}{Outcome, runs that answered} & & &\\
\cmidrule(lr){2-5}
Class ($N$) & SADE & CC-B. & ReAct & All & $\Delta$ & Over-d. & T.O.\\
\midrule
Correct-Fault (80) & .912 & .942 & .941 & .932 & --- & --- & 2\%\\
False-Premise (72) & .745 & .870 & .848 & .820 & $-.112$ & 17\% & 5\%\\
Wrong-Device (24) & .948 & .993 & .931 & .957 & $+.025$ & --- & 0\%\\
Wrong-Cause (24) & .901 & .981 & .850 & .911 & $-.021$ & --- & 0\%\\
\bottomrule
\end{tabular}
\vspace{-3mm}
\end{table}

\subsection{Accurate and Misleading Tickets (RQ1, RQ2)}
\label{sec:res-saturated}

We first measure the setting that existing benchmarks evaluate. That is a genuine fault accurately reported in the ticket. Table~\ref{tab:e1} reports the results.  On the 80 Correct-Fault scenarios, the three agents achieve average outcome scores of 0.912 (SADE), 0.942 (CC-Baseline), and 0.941 (ReAct), indicating that diagnosing a real fault from an accurate ticket is nearly solved, and further scenarios of this kind are less likely to separate one agent design from another. 


The Wrong-Device and Wrong-Cause scenarios ask whether a misleading ticket changes this (RQ2). The results show that it is not the case. The agents average 0.957 in the outcome score when the ticket incorrectly points to a healthy device and 0.911 when it incorrectly asserts a mechanism. Their proposed repairs remain correct in all three fault classes (average fix scores of 0.946, 0.919, and 0.897 as shown in Table~\ref{tab:fixreason}). The investigation traces explain this robustness. The fault produces symptoms that the agents can observe directly, such as a stopped daemon, a missing route, or a filtered port. Their probing contradicts those symptoms, and as a result, the agent discards the ticket's incorrect claim.

 

These results also mark the limit of what fault-based evaluation can measure: as long as a fault exists, an agent that verifies the ticket and an agent that blindly follows the ticket reach the same diagnosis, and these two behaviors cannot be separated. They separate only when the ticket reports a fault that does not exist


\subsection{False-Premise Tickets (RQ2)}
\label{sec:res-healthy}
 
The False-Premise scenarios test the opposite condition: the network is functioning correctly, and, as a result, the correct diagnosis is that no anomaly exists. As shown in Table~\ref{tab:e1}, across these 72 scenarios, the mean outcome score significantly drops to 0.820, compared to the Correct-Fault mean of 0.932.
 
However, here, the agents do not fail due to missing evidence. The traces show that they correctly disprove the reported symptom; they fail by continuing to probe until some genuine but benign standing network condition can be presented as the root cause, instead of reporting that nothing is wrong (i.e., over-diagnosis). Across the three passes, SADE reports a fault on the healthy network in 24\% of its answered runs (51/210), ReAct in 16\% (33/209), and CC-Baseline in 11\% (21/198) as presented in (Table~\ref{tab:endings}, Standard).


This ranking is the reverse of diagnostic sophistication: the skills layer that helps SADE localize genuine faults gives it more ways to find something to point at on a healthy network. The reasoning scores presented later confirm that the problem is not unmethodical investigations. SADE's traces are the best-grounded of the three agents on exactly this class (0.933, against 0.892 for CC-Baseline and 0.845 for ReAct; Table~\ref{tab:fixreason}).  What the agents lack is a criterion for concluding, with evidence, that no fault exists. Prior benchmarks cannot observe this failure because a fault is always present~\cite{nika2025}.
 

\begin{table}[t]
\caption{E2: outcome by persona on the 72 healthy networks}
\label{tab:e2}
\footnotesize
\setlength{\tabcolsep}{3.5pt}
\begin{tabular}{@{}l c cccc cc@{}}
\toprule
& & \multicolumn{4}{c}{Outcome, runs that answered} & &\\
\cmidrule(lr){3-6}
Persona & ID & SADE & CC-B. & ReAct & All & Over-d. & T.O.\\
\midrule
Standard & 3.2 & .745 & .870 & .848 & .820 & 17\% & 5\%\\
Naive & 2.1 & .773 & .911 & .789 & .820 & 16\% & 9\%\\
Novice & 1.9 & .643 & .793 & .609 & .675 & 31\% & 11\%\\
Novice-confident & 1.9 & .704 & .794 & .654 & .714 & 28\% & 13\%\\
Naive-unsure & 2.1 & .698 & .486 & .670 & .638 & 35\% & 19\%\\
Naive, no detail & 0.0 & .597 & .283 & .514 & .497 & 47\% & 22\%\\
\bottomrule
\end{tabular}
\vspace{-3mm}
\end{table}

\subsection{Reporting Personas (RQ3)}
\label{sec:res-personas}
 
Table~\ref{tab:e2} reports the average outcome score for each persona rewrite. All six rows run on the same 72 verified-healthy networks, so any difference among them is due to wording alone. The naive persona, which is non-technical but asserts a definite (and false) cause and names the component it blames, behaves similarly to the standard technical ticket. For the novice persona, which is equally non-technical but vague and offers no cause, the average outcome score falls to 0.675. The premises are equally false in both cases, since the network is healthy throughout; what separates them is that one reporter hands the agent a verifiable claim, and the other does not. Table~\ref{tab:e2} further shows the following :  \\ \vspace{-3mm}
 


\noindent{\bf Reporting tone (Novice vs.\ Novice-confident)}: Making the novice ticket confident rather than apologetic, with every identifier preserved, improves the outcome score slightly by $+0.039$. \\ \vspace{-3mm}


\noindent{\bf Including a definite cause (Naive vs.\ Naive-unsure)}: Withdrawing the naive reporter's certainty, so that the same suspected cause is only suggested rather than asserted and no repair is prescribed, drops the average outcome by 0.182 with the identifier count unchanged at 2.1, and raises over-diagnosis from 16\% to 35\%. \\ \vspace{-3mm}

\noindent{\bf Identifiers (Naive-unsure vs.\ Naive, no detail)}: Deleting the identifiers from the Naive-unsure ticket, with its content and tone otherwise fixed, drops the average outcome by a further 0.141 and raises over-diagnosis to 47\% for all the agents. \\ \vspace{-3mm}

Overall, the two control chains point in opposite directions. Adding confidence to the uncertain novice ticket (Novice vs.\ Novice-confident) makes the agents slightly more likely to reject the false report, the outcome rises by $+0.039$, and over-diagnosis falls from 31\% to 28\%. Adding uncertainty and then vagueness to the naive ticket (Naive, Naive-unsure, and No-detail rows) drops the average outcome from 0.820 to 0.638 and then to 0.497, and raises over-diagnosis from 16\% to 35\% and 47\%. Uncertainty in a wrong ticket, therefore, dominates the score far more than confidence does, the reverse of the expectation from work on sycophancy~\cite{sharma2024sycophancy}.  The agents can test and refute a definite claim against the live network, while a vague report leaves nothing to refute. While the class and persona averages answer the research questions, they conceal how the agents fail and the associated costs, which are investigated next.



\begin{table}[t]
\caption{Over-diagnosis and Turn budget exhaustion.}
\label{tab:endings}
\footnotesize
\setlength{\tabcolsep}{3.5pt}
\begin{tabular}{@{}l cc cc cc@{}}
\toprule
& \multicolumn{2}{c}{SADE} & \multicolumn{2}{c}{CC-B.} & \multicolumn{2}{c}{ReAct}\\
\cmidrule(lr){2-3}\cmidrule(lr){4-5}\cmidrule(lr){6-7}
Persona & Over-d. & T.O. & Over-d. & T.O. & Over-d. & T.O.\\
\midrule
Standard & 24\% & 3\% & 11\% & 8\% & 16\% & 3\%\\
Naive & 20\% & 4\% & 7\% & 17\% & 20\% & 6\%\\
Novice & 32\% & 6\% & 20\% & 21\% & 39\% & 6\%\\
Novice-confident & 27\% & 6\% & 19\% & 22\% & 37\% & 10\%\\
Naive-unsure & 28\% & 8\% & 51\% & 43\% & 32\% & 7\%\\
Naive, no detail & 34\% & 8\% & 74\% & 49\% & 45\% & 8\%\\
\bottomrule
\end{tabular}
\vspace{-3mm}
\end{table}

\begin{table}[t]
\vspace{-1mm}
\caption{Fix and reasoning scores by class and persona.}
\label{tab:fixreason}
\scriptsize
\setlength{\tabcolsep}{3pt}
\begin{tabular}{@{}l cccc cccc@{}}
\toprule
& \multicolumn{4}{c}{Fix} & \multicolumn{4}{c}{Reasoning}\\
\cmidrule(lr){2-5}\cmidrule(lr){6-9}
Class / persona & SADE & CC-B. & ReAct & All & SADE & CC-B. & ReAct & All\\
\midrule
Correct-Fault (80) & .952 & .962 & .924 & .946 & .937 & .917 & .821 & .891\\
False-Premise (72) & --- & --- & --- & --- & .933 & .892 & .845 & .890\\
Wrong-Device (24) & .903 & .979 & .873 & .919 & .944 & .957 & .868 & .923\\
Wrong-Cause (24) & .896 & .958 & .837 & .897 & .934 & .929 & .798 & .887\\
\addlinespace
Naive & --- & --- & --- & --- & .911 & .894 & .817 & .873\\
Novice & --- & --- & --- & --- & .891 & .845 & .760 & .831\\
Novice-confident & --- & --- & --- & --- & .920 & .865 & .767 & .851\\
Naive-unsure & --- & --- & --- & --- & .905 & .733 & .789 & .821\\
Naive, no detail & --- & --- & --- & --- & .877 & .688 & .718 & .774\\
\bottomrule
\end{tabular}
\vspace{-3mm}
\end{table}


\subsection{Per-Agent Failure Behavior}
\label{sec:res-signatures}



We next measure how each agent fails rather than how often. Table~\ref{tab:endings}
reports how a run on a healthy network ends and Table~\ref{tab:fixreason} the fix and
reasoning scores.

\noindent{\bf CC-Baseline depends most on the ticket}: it obtains the highest
outcome score of the three agents on the standard ticket (0.870) and the lowest on
the no-detail rewrite (0.283). Withdrawing only the reporter's certainty (Naive vs.\
Naive-unsure) drops its outcome by 0.425 and raises its over-diagnosis from 7\% to
51\%, and on the no-detail rewrite it exhausts the turn budget in 49\% of runs.
The wording of the ticket is therefore the decisive factor for this agent. \\
\noindent{\bf ReAct fails in the opposite way}: it submits a diagnosis in nearly
every run (timeouts of 3 to 10\%), but its verdicts are the least grounded. It has
the lowest reasoning score of the three agents on every fault class (0.798 to
0.868), the lowest fix score on every fault class, and the largest drop under
vague phrasing ($-0.239$ from the standard to the novice ticket). Its false
diagnoses name surface-level connectivity causes rather than the underlying
mechanism, consistent with the behavior NIKA reports~\cite{nika2025}. \\
\noindent{\bf SADE is the least affected by wording}: its outcome stays within
0.597 to 0.773 and its timeouts within 3 to 8\% across all six personas, and it
scores highest on the two most degraded ticket forms. The corresponding cost is a
false-alarm rate of 20 to 34\% on healthy networks that wording does not move, a
constant error that could in principle be calibrated away, unlike CC-Baseline's
collapse. These differences in how runs end also determine what they cost, which
we examine next. \\ \vspace{-3mm}

\subsection{Diagnostic Effort and Cost}
\label{sec:res-cost}

Cost decides whether an agent can be deployed, and Table~\ref{tab:effort} shows that effort follows the ticket rather than the fault. Scenarios with a real fault are the cheapest, at 13--18 tool calls per answered run, because the agent finds the fault, reports it, and stops. The vague rewrites are the most expensive: on the no-detail category CC-Baseline makes 30 tool calls against 13 on Correct-Fault and reads four times as many tokens per answered run (549k vs.\ 136k). Timeouts cost most of all: the 366 of them are 9\% of the 3{,}960 runs but consume 23\% of the US\$1{,}525 total spend. That burden falls unevenly, with 36\% of CC-Baseline's budget going to runs that produced no diagnosis against 10\% for SADE and 14\% for ReAct. The tickets that prior benchmarks omit are therefore the most expensive to process and the least likely to yield a diagnosis.




\begin{table}[t]
\vspace{-2mm}
\caption{Effort per agent: Tool calls and Cost.}
\label{tab:effort}
\footnotesize
\setlength{\tabcolsep}{4pt}
\begin{tabular}{@{}l cc cc c@{}}
\toprule
& \multicolumn{2}{c}{Tool calls} & \multicolumn{2}{c}{US\$} & \\
\cmidrule(lr){2-3}\cmidrule(lr){4-5}
Agent & answered & timeout & answered & timeout & Spend on t.o.\\
\midrule
SADE & 16.6 & 38.0 & 0.45 & 1.04 & 10\%\\
CC-B. & 22.1 & 49.2 & 0.43 & 1.07 & 36\%\\
ReAct & 19.5 & 45.5 & 0.12 & 0.38 & 14\%\\
\bottomrule
\end{tabular}
\vspace{-6mm}
\end{table}



\section{Conclusion}\label{sec:conclusion}

We presented FaulT-Bench, a benchmark that evaluates LLM troubleshooting agents on emulated networks using the kinds of tickets they will receive in operation, comprising 200 scenarios across eight network topologies, five of them translated from public practitioner labs. In 80 scenarios a genuine fault is described correctly, while in the remaining 120 the ticket reports a fault that does not exist, blames a healthy device, or asserts the wrong cause. The 72 false-premise tickets are further rewritten in five reporter personas over an unchanged network state. We also developed an evaluation harness that deploys, injects, verifies, and grades every run with a fixed LLM judge on three scores, with timeouts reported separately. Our evaluation yields two findings. When a fault is present, the agents localize it even from a ticket that blames the wrong device or asserts the wrong cause, because a genuine fault leaves evidence that systematic probing recovers. When no fault is present, the agents disprove the reported symptom but fail to conclude that nothing is wrong, continuing to probe until a benign condition can be promoted to a root cause. What governs the outcome is not the reporter's confidence but the absence of anything the agent can check against the network: removing the verifiable identifiers from a ticket costs 0.141 in outcome score, and the agents then fail in two distinct ways, SADE and ReAct by reporting a fault that is not there, and CC-Baseline by returning no verdict at all in up to half of its runs. FaulT-Bench therefore shows that treating the ticket as a hypothesis rather than an instruction is a capability distinct from diagnostic skill, and one that current agents lack. The principal limitation of our benchmark is that its tickets are authored rather than collected: real helpdesk reports are difficult to obtain for research, particularly at a scale that spans multiple topologies. The judge is itself an LLM, three agents under a single turn budget do not span the design space.

\newpage
\bibliographystyle{ACM-Reference-Format}
\bibliography{bib}










\end{document}